\documentstyle[11pt,IAUS215,twoside]{article}

\def\col#1{\empty} 
\def\bw#1{#1}       

\def\col#1{#1}      
\def\bw#1{\empty}

\def\edcomment#1{\iffalse\marginpar{\raggedright\sl#1\/}\else\relax\fi}
\reversemarginpar

\begin{document}
\vspace*{1.0cm}

\title{SCALING LAWS FOR DARK MATTER HALOS 
       IN LATE-TYPE AND DWARF SPHEROIDAL GALAXIES}

\author{John Kormendy}
\affil{Department of Astronomy, The University of Texas at Austin,
       1~University Station C1400, Austin, TX 78712-0259, USA}

\author{K.~C.~Freeman}
\affil{Research School of Astronomy and Astrophysics,
       Mount Stromlo Observatory, The Australian National University,
       Cotter Road, Weston~Creek, Canberra, ACT 72611, Australia}

\begin{abstract}
      Published mass models fitted to galaxy rotation curves are used to study 
the systematic properties of dark matter (DM) halos in late-type and dwarf 
spheroidal (dSph) galaxies. Halo parameters are derived by fitting non-singular
isothermals to $(V^2 - V_{\rm vis}^2)^{1/2}$, where $V(r)$ is the observed
rotation curve and $V_{\rm vis}$ is the rotation curve of the visible matter.
The latter is calculated from the surface brightness assuming 
that the mass-to-light ratio $M/L$ is constant with radius.  ``Maximum disk'' 
values of $M/L$ are adjusted to fit as much of the inner rotation curve as
possible without making the halo have a hollow core.  Rotation curve
decomposition becomes impossible fainter than absolute magnitude 
$M_B \simeq -14$, where $V$ becomes comparable to the velocity dispersion of 
the gas.  To increase the luminosity range further, we include 
dSph galaxies, which are physically related to spiral and irregular
galaxies. Combining the data, we find that DM halos satisfy well 
defined scaling laws analogous to the ``fundamental plane'' relations for
elliptical galaxies.  Halos in less luminous galaxies have smaller core radii
$r_c$, higher central densities $\rho_0$, and smaller central velocity
dispersions $\sigma$.  Scaling laws provide new and 
detailed constraints on the nature of DM and on galaxy formation and evolution.
Some simple implications include:

      1 -- A single, continuous physical sequence of increasing mass extends
from dSph galaxies with $M_B \simeq -7.6$ to Sc{\thinspace}I galaxies with
$M_B \simeq -22.4$.

      2 -- The high DM densities in dSph galaxies are normal for 
such tiny~galaxies.  Since virialized density depends on collapse redshift
 $z_{\rm coll}$, $\rho_0 \propto (1 + z_{\rm coll})^3$, the smallest dwarfs 
formed at least $\Delta z_{\rm coll} \simeq 7$ earlier than the biggest spirals.

      3 -- The high DM densities of dSphs implies that they are real galaxies 
formed from primordial density fluctuations.  They are not tidal fragments. 
Tidal dwarfs cannot retain even the low DM densities of their giant-galaxy 
progenitors.  In contrast, dSphs have higher DM densities than do giant-galaxy 
progenitors.

      4 -- The fact that, as luminosity decreases, dwarf galaxies become much
more numerous and also more nearly dominated by DM raises the possibility~that
there exists a large population of objects that are completely dark.  Such
objects are a canonical prediction of cold DM theory.  If they exist, ``empty 
halos'' are likely to be small and dense -- that is, darker versions of Draco
and UMi.

      5 -- The slopes of the DM parameter correlations provide a measure on
galactic mass scales of the slope $n$ of the power spectrum 
$|\delta_k|^2 \propto k^n$ of primordial density fluctuations.
Our preliminary results not yet corrected for baryonic compression of DM
give $n \simeq -1.9 \pm 0.2$.  This is consistent with cold DM theory.
\vspace*{-1cm} 
\end{abstract}

\section{Introduction}

\pretolerance=15000  \tolerance=15000

      This paper updates our derivation (Kormendy 1988, 1990; Kormendy \&
Freeman 1996) of scaling laws for DM halos of Sc -- Im and dwarf spheroidal
(dSph) galaxies.  We show that DM halos in less luminous galaxies have smaller 
core radii $r_c$, higher central densities $\rho_0$, and smaller central 
velocity dispersions~$\sigma$.  These scaling laws are analogous to the
fundamental plane relations for elliptical galaxies (Djorgovski \& Davis 1986,
1987; Faber et al.~1987; Dressler et al.~1987; Djorgovski, de Carvalho, \& Han 
1988; see Kormendy \& Djorgovski 1989 for a review), and they are interesting 
for the same reason: they provide new constraints on galaxy formation and 
evolution.  Simple conclusions are discussed in \S{\thinspace}4.  A detailed
discussion will be published in Kormendy \& Freeman (2003).

      Halo parameters for giant galaxies are derived by decomposing rotation
curves $V(r)$ into visible matter and DM contributions (van Albada et al.~1985).
At galaxy absolute magnitudes $M_B \gg -14$, rotation curve decomposition
becomes impossible as $V$ decreases (Tully \& Fisher 1977) and becomes
similar to the velocity dispersion of the gas. Pressure-supported galaxies
are not flat.  DM central densities can still be derived, e.{\thinspace}g., by
fitting King (1966) models to the density and velocity dispersion profiles of
dSph galaxies.  But DM $r_c$ and $\sigma$ can no longer be measured.  In this 
paper, we combine these data to investigate the systematic properties of DM
halos over a large range of galaxy luminosities.

      Only Sc -- Im and dwarf spheroidal (dSph) galaxies are included. Galaxies 
of type E -- Sbc are omitted for two reasons that result from their bulge 
components.  (1) Rotation curve decomposition must deal with two visible-matter
components that have different unknown mass-to-light ratios.  Therefore it is
less reliable.  (2) Gravitational compression of the DM by the baryons has
substantially modified the halo when the visible mass density is high.  Many 
Sa -- Sbc galaxies satisfy the DM correlations, but others deviate in the
direction of small $r_c$ and large $\rho_0$ (Kormendy 1988, 1990).  This is
consistent with baryonic compression.  Further evidence for baryonic compression
is presented in Athanassoula, Bosma, \& Papaioannou (1987, hereafter ABP).
Baryonic compression corrections are omitted here but will be included in
Kormendy \& Freeman (2003). 

\section{Core Parameters of DM Halos}

\subsection{Rotation Curve Decomposition.~I.~Technique}

      Consider first the rotation curve of an isothermal sphere in the ideal
case where we can measure a massless disk embedded in it.  Then at
$r \ll r_c$, \vskip -1pt
\begin{equation}
V \simeq \biggl({  {4\pi G\rho_0}\over{3}   }\biggr)^{1/2}~r,      
\end{equation}
and at $r \gg r_c$,
\begin{equation}
V \simeq 2^{1/2}\thinspace\sigma = 
      2^{1/2}~\biggl({ {4\pi G\rho_0r_c^2}\over{9} }\biggr)^{1/2}~,  
\end{equation}
where $G$ is the gravitational constant.  If we observe only the $V \propto r$
part of the rotation curve, we can measure $\rho_0$ but not $r_c$ or $\sigma$.  
Because of this, $\rho_0$ is often the only halo parameter that we can measure
in low-luminosity galaxies.  In contrast, if the measurements reach far enough
into the $V =$ constant part of the rotation curve, then all three parameters
can be measured.  Dwarf \hbox{Sc -- Im} galaxies come closest to the above
ideal, because visible matter contributes only a small fraction of the total
mass.

      More generally, visible matter dominates the central part of the 
rotation curve, and a multicomponent mass model is required.
The rotation curve of the visible matter is calculated from the
brightness distribution assuming that the mass-to-light ratio $M/L$ of
each component is constant with radius.  Values of $M/L$ are adjusted
to fit as much of the inner rotation curve as desired.  H{\thinspace}I gas
is taken into account separately.
Molecular gas is assumed to follow the light distribution, so it is
included in $M/L$.  Then, given the total rotation curve $V_{\rm vis}$
of the visible matter, the halo rotation curve is $V_{\rm DM}(r)=(V^2 -
V_{\rm vis}^2)^{1/2}$.  A model such as an isothermal is then fitted to
$V_{\rm DM}$ to derive the halo asymptotic velocity
$V_\infty=2^{1/2}\sigma$, $r_c$, and $\rho_0$.  Rotation curve decompositions
have now been published for $\sim 100$ galaxies.  Of these, 55 survive our
selection cuts (\S\thinspace2.3).

\subsection{Rotation Curve Decomposition.~II.~Our Assumptions}

      Many authors emphasize that mass modeling is uncertain (van Albada
et al. 1985; Skillman et al.~1987; Lake \& Feinswog 1989).  Lake and Feinswog 
point out that if measurement errors are interpreted strictly, few observations
of rotation curves reach large enough radii to determine halo parameters
uniquely.  Therefore the present results depend on the following assumptions:

      1 -- Rotation curves that flatten out to $V \simeq$ constant are assumed
to stay flat outside the radius range measured.  

      2 --  We use maximum disk decompositions.  This requires discussion. The
unknown mass-to-light ratio of the disk is a problem -- the ratio of
visible to dark mass can be varied greatly while preserving a good fit to
$V(r)$.  As the amount of visible mass is reduced, the central DM density 
must be increased and its core radius must be decreased.  The extreme models
(van Albada et al. 1985, Fig.~4 and 8) are usually the maximum disk mass that
does not require the halo to have a hollow core and a solution with $M/L = 0$.
In giant galaxies, these solutions are very different.  If we had no additional
constraints, we could say little about halo properties.  Fortunately, we have
other constraints.  We cannot let $M/L$ get arbitrarily small.  We observe
structures such as bars and spiral density waves that require disks to be 
self-gravitating.  ABP turned this qualitative remark into a practical constraint
on $M/L$ by applying Toomre's (1981) swing amplifier instability criterion and 
requiring that the disk have the proper density to give the observed spiral 
structure (i.{\thinspace}e., two arms but not one). The results converged on 
essentially the maximum disk solutions for 18 of 21 Sc\thinspace--{\thinspace}Im 
galaxies studied.  The resulting mass-to-light ratios imply plausible young stellar 
populations.  In general, some evidence favors maximum disks (e.{\thinspace}g., 
Taga \& Iye 1994; Sackett 1997; Debattista \& Sellwood 1998; Bosma 1999; Sellwood
\& Moore 1999; Weiner, 
Sellwood, \& Williams 2001; Gerhard 2004; Athanassoula 2004; Weiner 2004), and other
evidence suggests that some disks are submaximal (e.{\thinspace}g., Bottema 1993, 
1997; Courteau \& Rix 1999).  Our choice of maximum disk 
solutions affects only giant galaxies; dwarfs are so DM dominated that $M/L$
uncertainties have little effect.  If we used ``Bottema disks'' instead of
maximum disks, parameter correlations with galaxy luminosity would be shallower.

      3 -- Halo are assumed to have non-singular isothermal mass distributions.
IAU Symposium 220 focuses in part on the well known collision (Moore 1994)
between the prediction that cold DM (CDM) has cuspy central density profiles 
$\rho(r)$ (e.{\thinspace}g., Navarro, Frenk, \& White 1996, 1997 [NFW], who
find that $\rho \propto r^{-1}$) and observational evidence that at least dwarf
galaxies have flat cores.  
For the purposes of this paper, the difference between isothermals and NFW 
profiles is nontrivial but relatively benign.  An analogous problem arose with
elliptical galaxies: Early studies of cores, including the discovery of
fundamental plane correlations (Kormendy 1984; Lauer 1985; Kormendy
1987b,{\thinspace}c), were published  before {\it Hubble Space Telescope\/}
({\it HST\/}) showed that high-luminosity ellipticals have cuspy cores with
projected densities $\Sigma(r) \propto r^{-m}$, $m \simeq 0$ to 0.25 at small
radii (e.{\thinspace}g., Lauer et al.~1995).  But pre-{\it HST\/} observations
of core radii and central densities probe the same physics as (and, in fact, 
are roughly proportional to) {\it HST\/} measurements of profile break radii 
and densities (Kormendy et al.~1994).  Most results deduced from ground-based
photometry remain valid.  We expect that the present DM parameters will prove
to measure the relevant physics when the form of the halo density profile is
better known.  That is, we consider $r_c$ as an approximate profile break 
radius and $\rho_0$ as a measure of the density at $r_b$ or averaged inside
$r_b$; this should be valid whether halos are isothermal or not.

\subsection{Rotation Curve Decomposition.~III.~Galaxy Selection Criteria}

      We take DM parameters from published decompositions with as few changes 
as possible consistent with the assumption that halos are isothermal, with a 
uniform distance scale, and with the following selection criteria:

   1 -- Morphological types are restricted to Sc -- Im and dSph, as noted~above.
Late-type and spheroidal galaxies are physically related (Kormendy 1985,
1987c; Binggeli \& Cameron 1991; Ferguson \& Binggeli 1994).  Most dSph
companions of our Galaxy have had episodes of star formation in the past 1 -- 8
Gyr (Da~Costa 1994 and Mateo 1998 provide reviews); they presumably turned from 
irregulars into spheroidals since that time (Kormendy \& Bender 1994).  In fact,
the distinction between galaxy types has blurred as H{\thinspace}I gas has
been found in or near a few spheroidals (e.{\thinspace}g., Sculptor: Knapp
et al.~1978; Carignan et al.~1998; see Mateo 1998 for a review).  It seems
physically reasonable to include Sc -- Im and dSph galaxies in the same parameter 
correlation diagrams.

      2 --  We discard most galaxies with inclinations $i < 40^\circ$.  Broeils 
(1992) remarks that H{\thinspace}I rotation curve derivations
are less accurate when the galaxy is too face-on.  Also, there is some danger
that oval distortions (Bosma 1978; Kormendy 1982) result in incorrect estimates
of inclinations. Because they provide much-needed leverage at high luminosities,
we kept four $i < 40^\circ$ galaxies from ABP: M{\thinspace}101, NGC 5236, NGC
6946, and IC 342.  Inclination is not a critical selection criterion; most
nearly face-on galaxies satisfy the DM correlations.

      3 -- The most important selection cut is to ensure that rotation curves
reach large enough radii to constrain the DM parameters.  After some
experimentation, we decided to require that the rotation curve measurements
reach out to at least 4.5 exponential scale lengths of the disk.  The peak in
$V(r)$ for an exponential occurs at 2.2 scale lengths (Freeman 1970), so the
above choice ensures that the outer disk rotation curve drops significantly 
over the radius range in which we have velocity data.  Observing a flat 
rotation curve then provides good constraints on DM parameters.  The fussy
choice of the ratio 4.5 resulted from a desire to keep a few galaxies that
provide leverage at the high-luminosity end of the correlations.  The radius 
cut is not applied slavishly; we keep a few galaxies that slightly violate 
the above criterion (ratio 3.5 -- 4.4) but that are sufficiently halo dominated
that the DM parameters are well determined.  These galaxies are DDO 127, 
DDO 154, DDO 168, NGC 247, and IC 2574.  In general, the radius cut is
important; if we do not use it, we get a substantially larger galaxy sample,
and it mostly is consistent with the DM correlations, but it shows considerably
larger scatter than Figures 2 -- 4.

      Besides the selection cuts, we adopt the following procedures
to make parameters from different sources be as consistent with each other
as possible.

      Following Broeils (1992), we base distances on the Virgocentric flow model
of Kraan-Korteweg (1986).  However, the zeropoint is based on distances from
Cepheids and from surface brightness fluctuations (Ferrarese et al.~2000;
Tonry et al.~2001).  The distance to the Virgo Cluster is taken to be 
$D$ = 16.5 Mpc, corresponding to a Hubble constant of $H_0 = 70$ km s$^{-1}$
Mpc$^{-1}$.  The center of the Virgo Cluster is assumed to be at ($l$, $b$) = 
(281$^\circ$, 75$^\circ$) (Binggeli, Tammann, \& Sandage 1987).  The
Virgocentric infall velocity of the Local Group is assumed to be 220 km 
s$^{-1}$.  When accurate distances of our galaxies are known, e.{\thinspace}g.,
from Cepheids, they are adopted.  Most sources used essentially the above
distance scale; when an author did not, we assumed that $r_c \propto D$ and 
that $\rho_0 \propto D^{-2}$.  This is not strictly correct, because gas and
dynamical masses scale differently with distance.  But the errors are small on
the scales of Figures 2 -- 4.  

      Galactic absorption corrections are from Burstein \& Heiles (1984) or 
equivalently from the RC3 (de Vaucouleurs et al.~1991).  Absolute magnitudes
are corrected for internal absorption as in Tully \& Fouqu\'e (1985). 

\subsection{Matching Pseudo-Isothermal Models to the Isothermal Sphere}

      In carrying out rotation curve decompositions, some authors model the
DM with a nonsingular isothermal sphere; then we adopt the parameters
$r_c$, $\rho_0$, and $\sigma$ without modification.  Other authors use the
``pseudo-isothermal sphere'' (hereafter the PITS), i.{\thinspace}e., the
approximation that the volume density is
\begin{equation}
\rho\thinspace(r) = \rho_0/(1 + r^2/a^2)~.
\end{equation}
They then derive parameters $\rho_0$, $a$, and the asymptotic circular velocity
$V(\infty)$ as $r \rightarrow \infty$.  No compelling physical argument favors
one model over the other.  However, if we want to combine data
from different sources, we need to correct parameters determined using the PITS
to the ones that would have have been measured using the isothermal.  An
exact correction is not possible, because equation (3) is a poor approximation
to the isothermal sphere except as $r \rightarrow \infty$.  Figure 1 shows three
possible scalings of the rotation and velocity dispersion curves of the PITS to
the isothermal.  All three scalings are shown to emphasize the physical
difference between the two models.  The scalings are not equally plausible, but
each has a simple underlying motivation.  We will find that the best scaling is 
intermediate between the middle and right panels of Figure 1. 

\begin{figure}[ht!]
\centering
\vspace{9cm}

\includegraphics{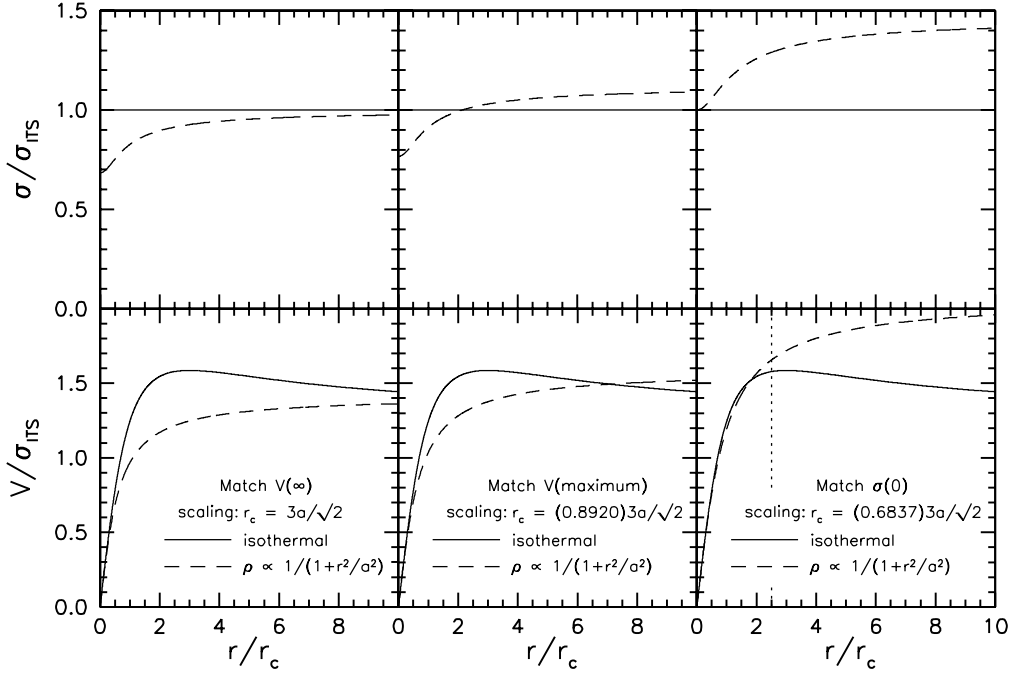}

\caption{Rotation curves ({\it bottom\/}) and velocity dispersion
profiles ({\it top\/}) for the isothermal and analytic halos normalized
by the velocity dispersion $\sigma_{\rm ITS}$ of the isothermal
sphere. The analytic halo is scaled in $r$ and $V$ so that both
models have the same central density and ({\it left\/}) asymptotic
rotation velocity, ({\it middle\/}) maximum rotation velocity, and
({\it right\/}) central velocity dispersion.  In the bottom-right
panel, the analytic and isothermal halos have similar rotation curves
out to $r/r_c \simeq 2.5$ ({\it vertical dotted line\/}), i.{\thinspace}e.,
over the radius range of typical HI rotation curves.  However, the two models
extrapolate very differently as $r \rightarrow \infty$.}
\label{fig1}
\end{figure} 

      In all three scalings, the central densities of the two models are the
same.  The panels differ in how radii and velocities are scaled.

      At left, the PITS is scaled in $r$ and $V$ so that the central densities
and asymptotic circular velocities are the same.  Then the core radius $r_c$ of
the isothermal is $r_c = 3a/\sqrt{2}$.  The dispersion scale in the top panel
is the velocity dispersion of the isothermal sphere, but this is equal
to the local dispersion of the PITS only as $r \rightarrow \infty$.  For most
galaxies, the H{\thinspace}I distribution does not extend beyond $\sim 2.5 r_c$.
 Here the {\it local\/} dispersion of the PITS is less than that of the
isothermal (Fig.~1, {\it top left\/}).  The reason is that an isotropic system 
with the density distribution of equation (3) is not isothermal; its velocity 
dispersion is 
\begin{equation}
  \sigma^2(r) = V^2(\infty){\thinspace}(1 + x^2) \biggl({{\pi^2}\over{8}} - 
                {{\tan^{-1} x}\over{x}} -
                {{(\tan^{-1} x)^2}\over{2}}\biggr){\thinspace}, 
\end{equation} 
where $x = r/a$.  Since the analytic model matches the isothermal badly in both
$V(r)$ and $\sigma(r)$, we discard this scaling.

      Given the limited extent of H{\thinspace}I data, it seems more realistic
to scale the PITS so that its rotation curve is similar to that
of the isothermal at $r \la 2.5{\thinspace}r_c$.  We get such a scaling
\hbox{(Fig.~1, {\it right\/})} if we solve our problem with the halo dispersion
by making  the velocity dispersion of the corresponding isothermal be equal to
the {\it central\/} dispersion \hbox{$\sigma(0) = 
(\pi^2/8 - 1)^{1/2}{\thinspace}V(\infty) = 0.4834{\thinspace}V(\infty)$} 
of the PITS. Then $r_c = (0.6837)3a/\sqrt{2} = 1.4503{\thinspace}a$ and $\sigma 
= 0.4834{\thinspace}V(\infty)$.  The resulting parameters are related in the
normal way for an isothermal sphere, $\sigma^2 = 4 \pi G \rho_0 r_c^2 / 9$.

      The virtue of this scaling is that the two halo rotation curves are
similar over the radius range in which they are actually fitted to the data
(Fig.~1, {\it bottom right\/}).  However, there is a problem (Fig.~1, {\it
top right\/}).  This scaling forces $V(\infty)$ to be much larger than the
velocity of the flat part of the rotation curve.  An examination of published
decompositions shows that authors almost never extrapolate implicitly to such a
large maximum rotation velocity -- one that is never observed.  The exceptions
are some rotation curve decompositions in which $V_{\rm DM} \propto r$ over the
whole radius range of the data, i.{\thinspace}e., cases in which $r_c$ is
completely unconstrained.  Examples are the decompositions of NGC 7331 and NGC
6674 in Broeils (1992).  We discard these decompositions anyway.

      This suggests a compromise scaling like the one in the middle panel.  Here
$V(\infty)$ is scaled to the {\it maximum\/}, not the asymptotic, rotation
velocity of the isothermal.  Then, when authors use equation (3) to derive
$\rho_0$, $a$, and $V(\infty)$ via rotation curve decomposition, we would adopt
$r_c = 1.8918{\thinspace}a$, the authors' quoted value of $\rho_0$, and a halo
velocity dispersion $\sigma = 0.6306{\thinspace}V(\infty)$.  Again, these
parameters are related as normal for an isothermal sphere.

      In practice, there is no guarantee that any of the above scalings 
represents what happens when one author uses the PITS and another uses the
isothermal in rotation curve decomposition.  We note again that Figure 1 is
included to emphasize that the PITS is not very isothermal.  However, we will
find in \S\thinspace3 that enough decompositions have been published using 
each of the above models so that we can derive the DM parameter correlations
separately for each model. We will then derive the best scaling of one model to
the other by matching their respective DM correlations.  This scaling is used
to construct Figure 4, which combines data from all sources.  It is 
intermediate between the middle and right-hand panels of Figure 1,
i.{\thinspace}e., 
$\rho_0 = 0.9255{\thinspace}\rho_{0,\rm PITS}$; 
$r_c = 1.6154{\thinspace}a$,
and 
$\sigma = 0.7334{\thinspace}\sigma_{\rm PITS} 
        = 0.5186{\thinspace}V_{\infty,\rm PITS}$.

\subsection{Central Densities of Dwarf Spheroidal and Irregular Galaxies}


\lineskip=0pt \lineskiplimit=0pt

       The smallest dSph galaxies allow us to greatly increase our leverage on
correlations of DM parameters with galaxy luminosity.  Note that only the halo
central density can be measured. Pioneering observations by Aaronson (1983) and 
by Aaronson \& Olszewski (1987) showed that the stellar velocity dispersions in
Draco and UMi are $\sim 10$ km s$^{-1}$, much larger than expected if the
galaxies are in equilibrium and if they consist only of old, metal-poor stars
($M/L_V$~$\la$~2.5).  High dispersions imply mass-to-light ratios
$M/L_V \sim 10^2$.  Moreover, the central DM densities $\rho_0 \sim 0.6$ to 1
$M_\odot$ pc$^{-3}$ are ``shockingly high. \dots~Indeed, these are the highest
central DM densities seen in any galaxy so far'' (Kormendy 1987a).  The latter
result was an early sign of the correlations in Figures 2 -- 4.  

      DM in dSph galaxies has important implications, so considerable
effort has gone into trying to find some escape.  Early worries included 
small-number statistics, measurement errors, atmospheric velocity jitter in the 
most luminous AGB stars, and especially unrecognized binary stars.  These have 
largely been laid to rest as more and fainter stars have been measured, as the
time baseline on measurements of individual stars has increased beyond a decade,
and as different authors have proved to agree (e.{\thinspace}g., Armandroff, 
Olswezski, \& Pryor 1995; Olszewski, Pryor, \& Armandroff 1996, see Tremaine 1987; 
Pryor 1992; Mateo 1994, 1997, 1998 for reviews).  As the number of dSph galaxies 
with dispersion measurements has increased (above reviews and Mateo et
al.~1998; Cook et al.~1999; C\^ot\'e et al.~1999; and Gallart et al.~2001),
escape routes that depend on rare events have become implausible. These include
the suggestion (Kuhn \& Miller 1989; Kuhn 1993) that the stars formerly in dSph
galaxies are unbound because of Galactic tides, so we overestimate the
masses of systems that are far from equilibrium.  The required orbital 
resonance works best if the dispersion is only marginally larger than the
escape velocity and if not too many systems need special engineering.  But
$M/L$ ratios of 10 -- 100 (not~2.5!) imply velocities that are inflated
by factors of $\sim 2 - 6$.  Piatek \& Pryor (1995), Oh, Lin, \& Aarseth (1995), 
Sellwood \& Pryor (1998), Klessen, Grebel, \& Harbeck (2003), and Wilkinson (2004)
argue convincingly that tides do not inflate velocity dispersions this much, 
especially not without producing velocity gradients across the galaxies that would 
have been seen.  This remains true even though apparently extratidal stars have 
been seen in some dSphs (e.{\thinspace}g., Irwin \& Hatzidimitriou 1995; 
Piatek et al. 2001, 2002; Palma et al.~2003).  In addition, some dSph galaxies are 
too far from our Galaxy to be affected by tides (Mateo 1998).  All nine dSph 
galaxies with dynamical analyses appear DM dominated.  It has become difficult to 
argue that we get fooled by special circumstances.

      One DM alternative is not addressed by the above arguments: Modified
Newtonian Dynamics (Milgrom 1983a, b, c; Milgrom \& Bekenstein 1987).  MOND 
has been much debated in recent years and will be revisited during this conference.
Thirty years of failure to identify all constituents of DM persuade us to treat
even exotic alternatives with respect.  However, while the jury is out, we
assume conventional gravity and treat measurements of high velocity dispersions
in dSph galaxies as detections of DM. 

      Accordingly, Figures 2 and 4 include DM central densities for all galaxies
with $\rho_0$ tabulated in Mateo (1998) plus And II (C\^ot\'e et al.~1999).
This includes four dI galaxies with virial measurements of $\rho_0$; three of
these are based on H{\thinspace}I and one (LGS3) is based on stellar dynamics
(Cook et al.~1999).  The central mass density has been corrected for visible
matter by subtracting 2.5 times the central, $V$-band volume brightness.
For all galaxies except Fornax, the correction is small.  In Fornax 
($M_B = -12.6$), visible matter accounts for approximately half of the central
density.  For this galaxy, $\rho_0$ is very uncertain.

      The ``error bars'' on $\rho_0$ require discussion.  The plotted densities
are derived by assuming that $r_c$ is the same for the visible and dark matter.
If $r_{c,\rm DM} \gg r_{c,\rm vis}$ as in the other galaxies in the figures,
then $\rho_0$ is smaller by a factor of 0.46 (Pryor \& Kormendy 1990; Lake
1990).  Anisotropic velocity dispersions reduce $\rho_0$ estimates by as much 
as an order of magnitude over the isotropic case (Pryor \& Kormendy 1990; Mateo
et al.~1993), although the extreme models are not good fits to the data.
Finally, model-independent lower limits to the DM density (Merritt 1987) are
typically $\rho_{0,\rm min} \simeq 10^{-2.2 \pm 0.17}$ $M_\odot$ pc$^{-3}$.
These limits are mostly too weak to be interesting.  However, for Leo II, UMi,
and Draco, they are $\log {\rho_{0,\rm min}/(1~M_\odot~{\rm pc}^{-3})} = -1.55$,
$-1.38$, and $-1.26$, respectively.  Even these low values are reasonably
consistent with the extrapolation of the DM correlations to low luminosities.

\section{DM Halo Scaling Laws}

      The correlations between halo $r_c$, $\rho_0$, and $\sigma$ and galaxy
absolute magnitude $M_B$ are illustrated in Figures 2 -- 4.  Figure 2 shows
results derived using isothermal halos.  Figure 3 shows results derived using
pseudo-isothermal halos.  Figure 4 combines all of the data, as discussed below.  

      The main result of this paper is that DM satisfies well defined scaling 
laws.  Halos in less luminous galaxies have smaller core radii, higher central
densities, and smaller central velocity dispersions.

      The scatter about the correlations is slightly smaller in Figure 2 than 
in Figure 3.  It is not surprising that decompositions based on the
isothermal sphere are better behaved, given the slow rise of the PITS rotation
curve to its asymptotic value (Fig.~1).  However, rotation curve decompositions
based on isothermal and PITS models separately give essentially the same
correlations.

      There are 31 galaxies with isothermal DM decompositions and 37 with PITS
decompositions.  Only 13 galaxies are common to both samples.  So the results 
in Figures 2 and 3 are largely independent.  Clearly, we want to combine 
the two samples.  We do this by scaling the correlations in Figure 3 to those 
in Figure 2.

      Least-squares fits to the correlations in Figure 2 (omitting NGC 4605) give:
\begin{eqnarray}
  \log{\rho_0}   &=&     -1.0383\,\log{r_c}    \quad~~\;\;- 1.0173~~~~({\rm rms}  = 0.17~{\rm dex})\;; \\
  \log{\sigma}\; &=& ~\;\,0.4812\,\log{r_c}    \quad~~\;\;+ 1.3808~~~~({\rm rms}  = 0.08~{\rm dex})\;; \\
  \log{\rho_0}   &=&     -1.2334\,\log{\sigma} \quad~~~~\,+ 0.2960~~~\;({\rm rms} = 0.31~{\rm dex})\;; \\
  \log{\rho_0}   &=& ~\;\,0.1130\,(M_B + 18)              - 1.9140~~~~({\rm rms}  = 0.28~{\rm dex})\;; \\
\,\log{r_c}\,    &=&     -0.1266\,(M_B + 18)              + 0.8615~~~~({\rm rms}  = 0.16~{\rm dex})\;; \\
  \log{\sigma}~  &=&     -0.0702\,(M_B + 18)              + 1.7942~~~~({\rm rms}  = 0.08~{\rm dex})\;. 
\end{eqnarray}
 
      Least-squares fits to the correlations in Figure 3 give:
\begin{eqnarray}
  \log{\rho_0}   &=&     -1.2045\,\log{r_c}    \quad~~\;\;- 1.1002~~~~({\rm rms}  = 0.16~{\rm dex})\;; \\
  \log{\sigma}\; &=& ~\;\,0.3987\,\log{r_c}    \quad~~\;\;+ 1.6653~~~~({\rm rms}  = 0.08~{\rm dex})\;; \\
  \log{\rho_0}   &=&     -2.1009\,\log{\sigma} \quad~~~~\,+ 2.1690~~~\;({\rm rms} = 0.36~{\rm dex})\;; \\
  \log{\rho_0}   &=& ~\;\,0.1575\,(M_B + 18)              - 1.8804~~~~({\rm rms}  = 0.43~{\rm dex})\;; \\
\,\log{r_c}\,    &=&     -0.1622\,(M_B + 18)              + 0.6532~~~~({\rm rms}  = 0.28~{\rm dex})\;; \\
  \log{\sigma}~  &=&     -0.0833\,(M_B + 18)              + 1.9289~~~~({\rm rms}  = 0.09~{\rm dex})\;. 
\end{eqnarray}

      The two samples have essentially the same average absolute magnitude: 
${<}M_B{>} = -18.12$ for objects analyzed with isothermals and 
${<}M_B{>} = -17.83$ for those analyzed with pseudo-isothermals.
The average of the above values is $M_B = -17.97$.  Requiring that the
correlations agree at $M_B = -18$ provides the scaling of the
PITS results to those measured with the isothermal sphere (ITS):
\begin{eqnarray}
  \rho_{0,\rm ITS} &=& 0.9255\:\rho_{0,\rm PITS}\;; \\
  r_{c,\rm ITS}    &=& 1.6154\:r_{c,\rm PITS}\;;     \\
  \sigma_{\rm ITS} &=& 0.7334\:\sigma_{\rm PITS} = 0.5186\:V_{\infty,\rm PITS}\;.
\end{eqnarray}
Here we explicitly identify the parameters derived with the different DM models.
This scaling is intermediate between the middle and right panels of Figure 1.

\vfill\eject

\begin{figure}[hb!]
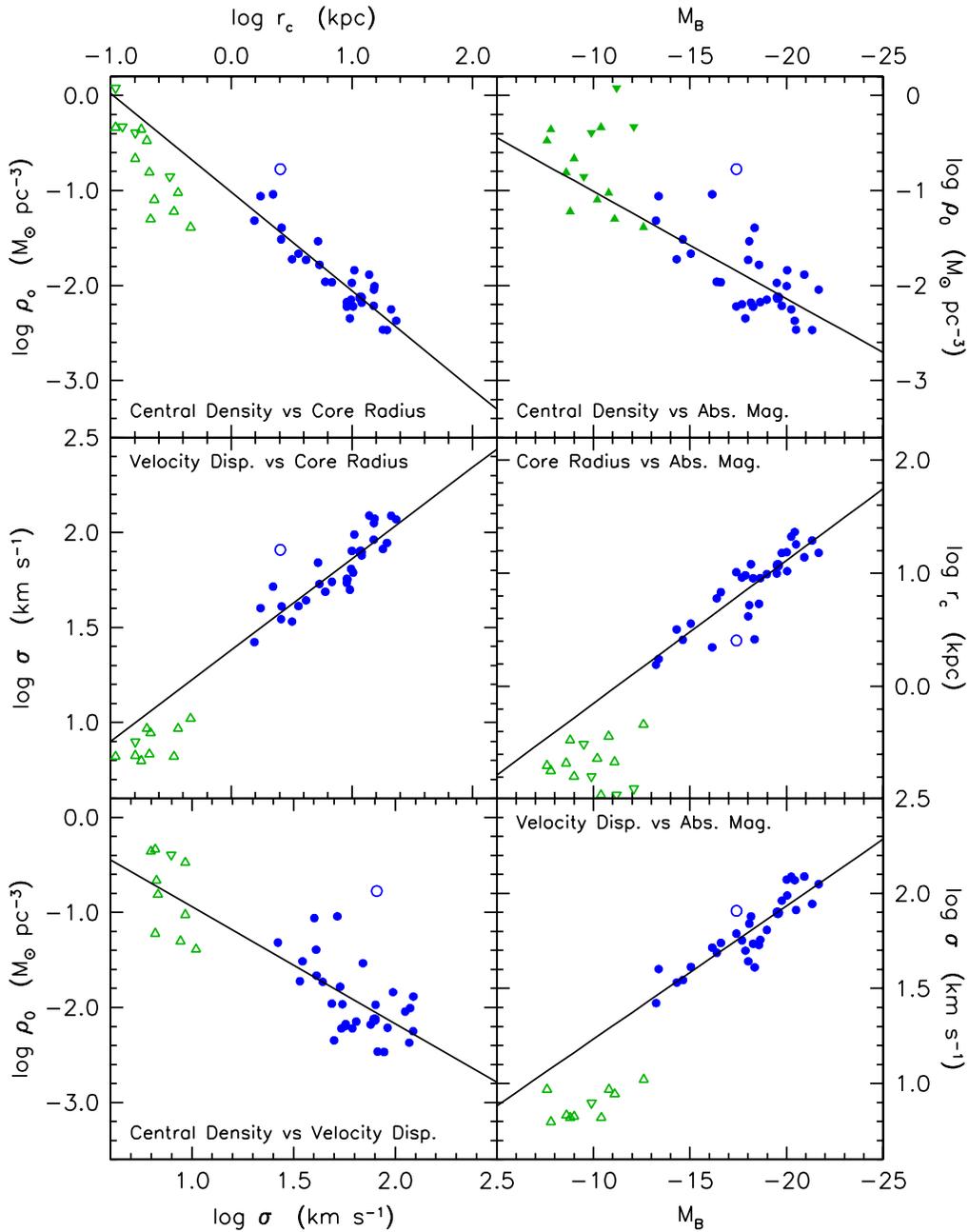

\centering
\vspace{17.0cm}

\bw{ \includegraphics{iau220iso.ps}}
\col{\includegraphics{iau220iso.cps}}

      \caption{Dark matter parameter correlations derived using isothermal 
halos in rotation curve decompositions of Sc -- Im galaxies ({\it circles\/}).
Central dark matter densities of dSph galaxies ({\it filled triangles}) and 
dI galaxies ({\it upside-down filled triangles\/}) are derived via King (1966) 
model fits; corresponding $r_c$ and $\sigma$ values are for the stars 
({\it open triangles\/}).   The lines are least-squares fits to the
\hbox{Sc{\thinspace}--{\thinspace}Im} galaxies omitting NGC 4605 ({\it open 
circle\/}).  Parameter sources:
  ABP;
  Blais-Ouellette et al.~(1999);
  Bosma et al.~(2003);
  Carignan \& Puche (1990);
  Carignan \& Purton (1998);
  Carignan et al.~(1988);
  C\^ot\'e et al.~(2000);
  Jobin \& Carignan (1990);
  Martimbeau et al.~(1994);
  Meurer et al.~1996);
  Puche et al.~(1991);
  Sicotte \& Carignan (1997);
  Verdes-Montenegro et al.~(1997).}
\label{fig2}
\end{figure}

\begin{figure}[hb!]
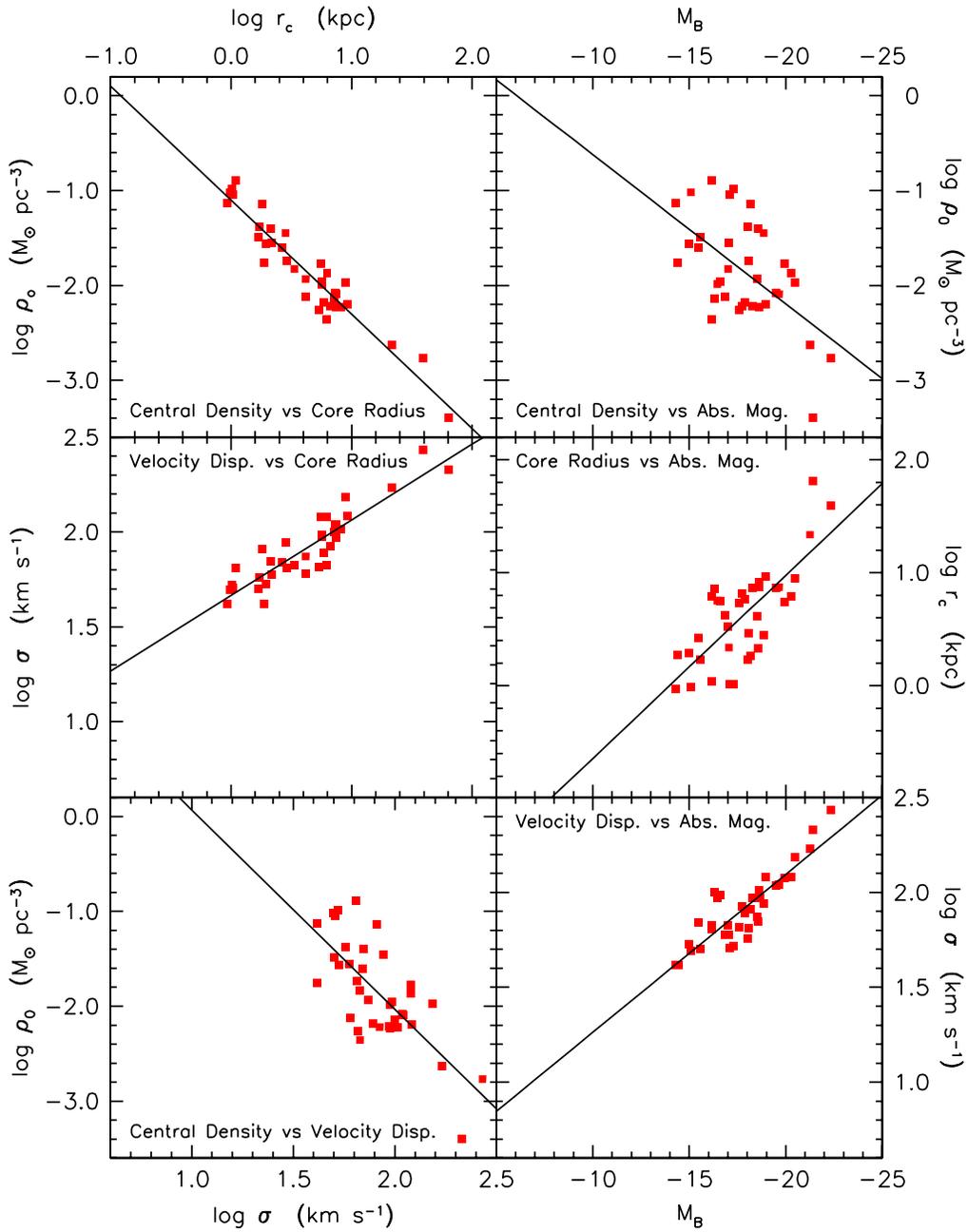

\centering
\vspace{17.0cm}

\bw{ \includegraphics{iau220pits.ps}}
\col{\includegraphics{iau220pits.cps}}

      \caption{Dark matter parameter correlations derived using the
pseudo-isothermal sphere to model DM halos in rotation curve decompositions of
Sc -- Im galaxies ({\it filled squares\/}).  The straight lines are least-squares
fits to all the points.  Parameter sources:
  Broeils (1992);
  Corbelli (2003);
  de Blok \& McGaugh (1996, 1997);
  de Blok, McGaugh, \& Rubin (2001);
  Meurer et al.~(1996);
  Meurer, Staveley-Smith, \& Killeen (1998);
  Miller \& Rubin (1995);
  Skillman et al.~(1987);
  Swaters, Madore, \& Trewhella (2000);
  Swaters et al.~(2003);
  van Zee et al.~(1997);
  Verheijen (1997);
  Weldrake, de Blok, \& Walter (2003).
}
\label{fig3}
\end{figure}

\eject\clearpage

\begin{figure}[hb!]
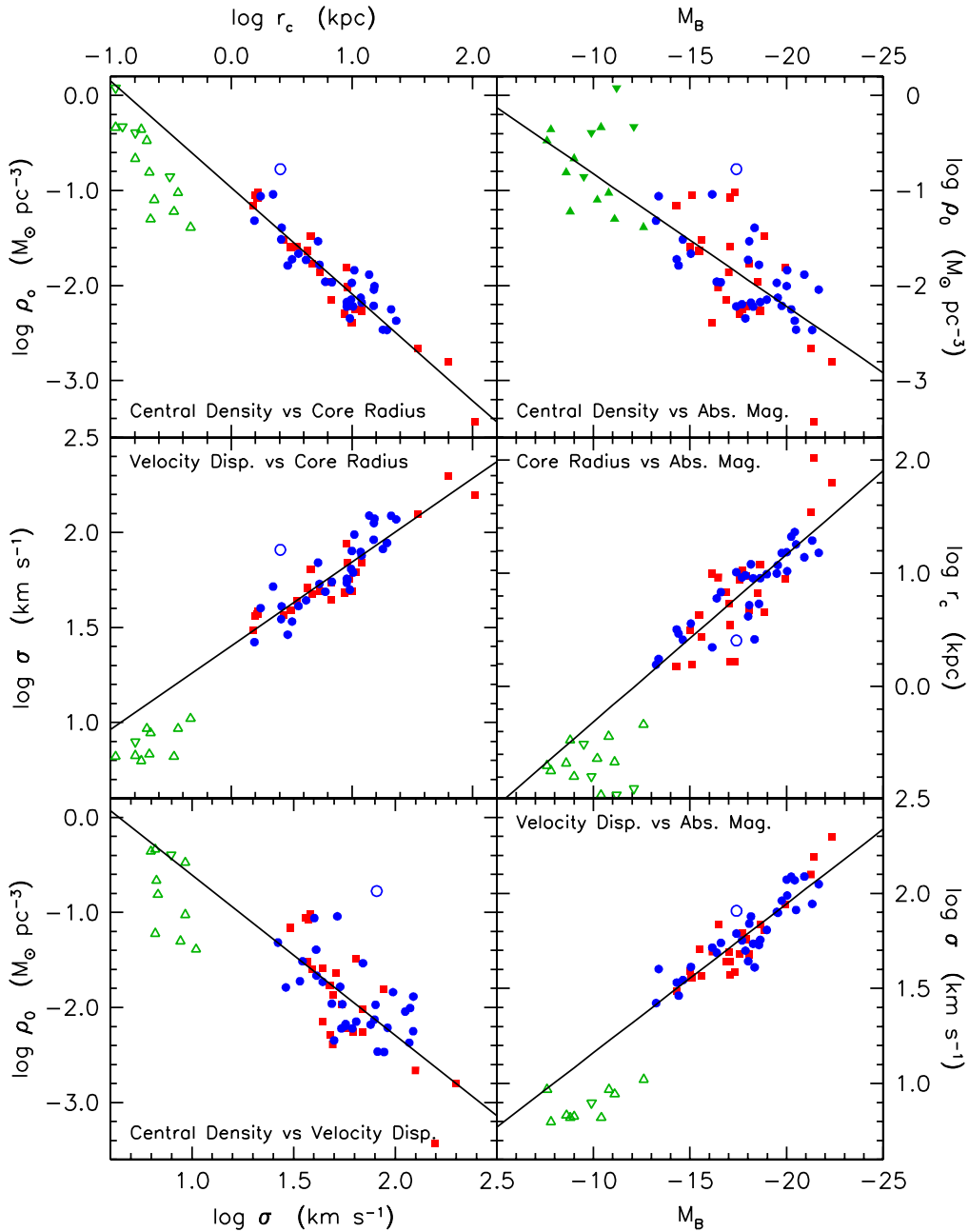

\centering
\vspace{17.0cm}

\bw{ \includegraphics{iau220all.ps}}
\col{\includegraphics{iau220all.cps}}

\caption{Dark matter parameter correlations for Sc{\thinspace}--{\thinspace}Im, 
dwarf spheroidal (dSph), and dI galaxies.  The symbols are the same as in 
Figures 2 and 3.  For dSph and dI galaxies, $\rho_0$ is a meaure of the DM 
but $r_c$ and $\sigma$ are visible-matter parameters.  The straight lines are
least-squares fits to the Sc -- Im galaxies omitting NGC 4605 ({\it open circle\/}).
NGC 4605 deviates from the correlations because the mass-to-light ratio of 
the disk was reduced significantly from its maximum disk value to kill off 
one-armed spiral instabilities (see ABP).  It is the only such galaxy retained
in the sample.  In Figures 2 -- 4, distances are based on a Virgocentric flow 
field calibrated using Cepheids and surface brightness fluctuations.
They correspond to $H_0 \simeq 70$ km s$^{-1}$~Mpc$^{-1}$.}
\label{fig4}
\end{figure}

\eject\clearpage

      Figure 4 shows the combined sample after scaling of the 
pseudo-isothermal results to those determined with isothermal halos.  
Based on 55 galaxies spanning 9 magnitudes in $M_B$, the correlations are 
quite robust.  Uncertainties in rotation curve decomposition are unlikely 
to threaten them, although the slopes are still uncertain.  If we keep 
only the galaxies with the most leverage on the DM, say the 11 galaxies 
whose rotation curves reach out to at least 10 times the scale length of 
the disk (one of these is NGC 3198, in which the ratio is 11.4),
then we get essentially the same correlations with essentially the same 
scatter.  Least-squares fits to the points in Figure 4 give:
 \begin{eqnarray}
  \log{\rho_0}   &=&     -1.1224\,\log{r_c}    \quad~~\;\;- 0.9692~~~~({\rm rms}  = 0.17~{\rm dex})\;; \\
  \log{\sigma}\; &=& ~\;\,0.4405\,\log{r_c}    \quad~~\;\;+ 1.4038~~~~({\rm rms}  = 0.08~{\rm dex})\;; \\
  \log{\rho_0}   &=&     -1.6902\,\log{\sigma} \quad~~~~\,+ 1.0838~~~\;({\rm rms} = 0.35~{\rm dex})\;; \\
  \log{\rho_0}   &=& ~\;\,0.1395\,(M_B + 18)              - 1.9410~~~~({\rm rms}  = 0.36~{\rm dex})\;; \\
\,\log{r_c}\,    &=&     -0.1481\,(M_B + 18)              + 0.8695~~~~({\rm rms}  = 0.22~{\rm dex})\;; \\
  \log{\sigma}~  &=&     -0.0784\,(M_B + 18)              + 1.7889~~~~({\rm rms}  = 0.09~{\rm dex})\;. 
\end{eqnarray}

      In physically more transparent terms ($10^9\:L_{B\odot}$ is $M_B = -17.03$),
\begin{eqnarray}
  \rho_0 &=& 0.0156~M_{\odot}~{\rm pc}^{-3}~~\biggl({L_B \over 
  10^9~L_{B\odot}}\biggr)^{-0.35};                                 \\
  r_c &=& 5.3~{\rm kpc}~~\biggl({L_B \over 
      10^9~L_{B\odot}}\biggr)^{0.37};                              \\
  \sigma &=& 52~{\rm km~s}^{-1}~~\biggl({L_B \over 
       10^9~L_{B\odot}}\biggr)^{0.20};~{\rm i.{\thinspace}e.,}     \\
  L_B &\propto& \sigma^{5.1};                                      \\
\vspace*{0.2cm}
  \rho_0 &=& 0.107~M_{\odot}~{\rm pc}^{-3}~~\biggl({r_c \over 
     1~{\rm kpc}}\biggr)^{-1.12};                                  \\
  \rho_0 &=& 0.0051~M_{\odot}~{\rm pc}^{-3}~~\biggl({\sigma \over 
100~{\rm km~s}^{-1}}\biggr)^{-1.69};                               \\
  \sigma &=& 25~{\rm km~s}^{-1}~~\biggl({r_c \over 
       1~{\rm kpc}}\biggr)^{0.44}\;.
\end{eqnarray}

      Equations 26 and 27 imply that halo surface density, which is
proportional to $\rho_0\,r_c$, is nearly independent of galaxy luminosity. 
This near-independence is illustrated explicitly in Figure 5.  It implies a 
Tully-Fisher (1977) relation similar to equation 29 (cf.~Fall 2002).  Since
mass $M \propto \rho_0 r_c^3$ and since $\rho_0 r_c \simeq$ constant,
$M^{1/2} \propto r_c$.  Then $M \propto \sigma^2 r_c$ implies that $M
\propto \sigma^4$.  We find that $L_B \propto \sigma^{5.1}$.  This
suggests that $M/L_B \propto L_B^{-0.22}$.  If we correct for the slight
dependence of halo surface density on $L_B$ (Fig.~5), then this becomes
$M/L_B \propto L_B^{-0.24}$.  Slope uncertainties 
are significant, so we should be cautious in how much we interpret
this result.  However, it is a plausible estimate of the degree to which 
small galaxies are more DM dominated than large ones.  UGC 2885 
($M_B = -22.3$) is brighter than UMi ($M_B = -7.6$) by 14.7 mag.
Over 15 mag, $M/L_B \propto L_B^{-0.24}$ implies that $M/L_B$ 
changes by a factor of 26.  This is entirely reasonable.  


\vfill\eject

\centerline{\null}

\begin{figure}[ht!]
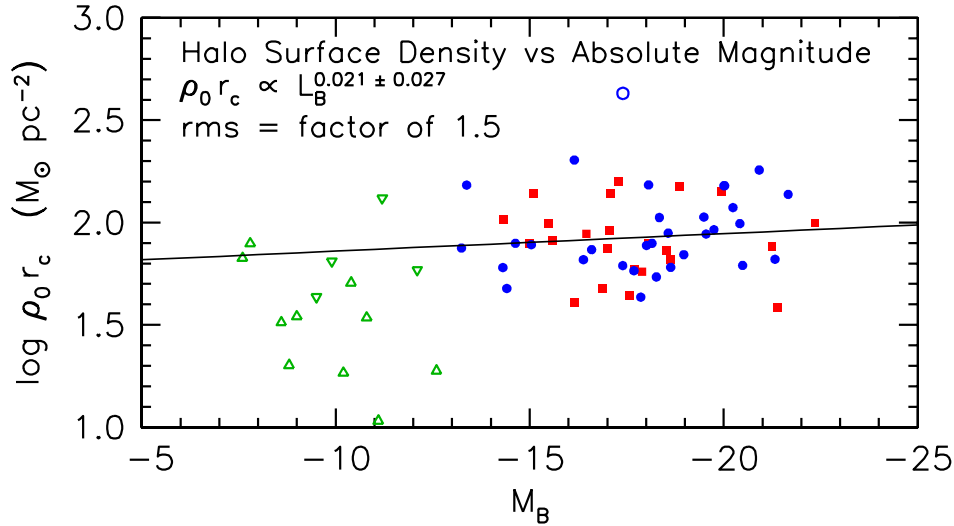

\centering
\vspace{6.2cm}

\bw{ \includegraphics{iau220sd.ps}}
\col{\includegraphics{iau220sd.cps}}

\caption{Logarithm of $\rho_0 r_c$, which is proportional to the projected
surface density of DM halos, as a function of absolute magnitude.  The symbols
are the same as in Figures 2 -- 4.  The straight line ({\it key\/}) is a 
least-squares fit to the \hbox{Sc -- Im} galaxies omitting NGC 4605.  Within
the errors, the surface densities of DM halos of late-type galaxies are
independent of galaxy luminosity.}
\label{fig4}
\end{figure}

\centerline{\null}
\vskip -35pt
\centerline{\null}

\section{Conclusions}

      Scaling laws are new constraints on the nature of DM and on 
galaxy formation and evolution.  Most of these remain to be explored. 
Simple implications include:

      1 -- The surprisingly high DM densities in dwarf spheroidals are
normal for galaxies of such low luminosity.  This implies that dSphs
are real galaxies and not tidal fragments.  Tides almost
certainly pull bound fragments out of more luminous galaxy progenitors, 
but they cannot retain even the relatively low DM densities in those
progenitors (Barnes \& Hernquist 1992), much less increase the DM density
to the high values characteristic of dwarf spheroidal galaxies.

      2 -- Dwarf spheroidal galaxies are not included in the
least-squares fits in Figures 2 and 4 because only $\rho_0$ can
be derived for their halos.  However, these $\rho_0$ values lie on the
extrapolation to low luminosity of the correlations for spiral and
irregular galaxies.  That is, the DM halos of dSph and Sc -- Im
galaxies appear to form a single physical sequence as a function of
DM core mass. 

      3 -- Since virialized density depends on collapse redshift
$z_{\rm coll}$, $\rho_0 \propto (1 + z_{\rm coll})^3$, the smallest 
dwarfs formed at least $\Delta\,z_{\rm coll} \simeq 7$ earlier than 
the biggest spirals.  Correction for baryonic DM compression will 
make $\rho_0$ smaller for giant galaxies.  This will slightly 
increase $\Delta\,z_{\rm coll}$.

      4 -- The visible matter parameters $r_c$ and
$\sigma$ of dSphs are a factor of about 2 smaller than their
extrapolated DM parameters.  This is reasonably consistent with the 
hypothesis that extreme dSphs have low visible matter densities 
($M/L_B \sim 10^2$) because they lost most of their baryons early.  Possible 
reasons include galactic winds (e.{\thinspace}g., Dekel \& Silk 1986) or 
the difficulty of holding onto baryons in shallow 
DM potential wells when the Universe was ionized
(e.{\thinspace}g., Klypin et al.~1999).  In the absence of a dark halo,
the loss of most baryons would unbind the few stars that had already
formed.  But since these galaxies contain DM halos, we expect instead
that the distribution of stars has expanded to fill the halo's core.
Unlike the situation in giant galaxies, visible matter and DM would
then have similar scale parameters.

      5 -- The fact that, as luminosity decreases, dwarf galaxies become much
more numerous and also more nearly dominated by DM raises the possibility that
there exists a large population of objects that are completely dark (Freeman
1987; Kormendy 1990; see also Tully 2004).  Undiscovered DM dwarfs would help to
solve the well known problem that the spectrum of initial density fluctuations 
predicted by CDM theory predicts far too many dwarf satellites of giant galaxies
(Moore et al.~1999; Klypin et al.~1999).  The favored explanation for why
these dwarfs are not seen is that they virialized early, before or during
the reionization of the Universe, and therefore lost or never captured
the canonical fraction of baryons because those baryons were too hot to be
confined in the puny potential wells of the dark dwarfs.  Our observations 
suggest that empty halos -- if they exist -- are likely to be small and dense
and to have small total masses. They would be darker versions of Draco and UMi.

      6 -- Djorgovski (1992) has compared an earlier version of the DM parameter 
correlations to the scaling laws predicted by hierarchical clustering (Peebles 
1974; Gott \& Rees 1975).  For a power spectrum of initial density fluctuations 
that is a power law in wavenumber $k$, $|\delta_k|^2 \propto k^n$, the size $R$,
density $\rho$, and velocity dispersion $\sigma$ of a bound object are related
approximately by
\begin{eqnarray}
  \rho   &\propto& R^{-3(3+n)/(5+n)}\,;   \\
  \rho   &\propto& \sigma^{-6(3+n)/(1-n)}\,;  \\
  \sigma &\propto& R^{(1-n)/(10+2n)}\,.
\end{eqnarray}
Here we have used the relation $\rho \propto \sigma^2 R^{-2}$ for an isothermal
sphere.  Djorgovski pointed out that the DM parameter correlations in Kormendy
(1990) imply that $n \simeq -2.45$, close to the value $n \simeq -2$ expected
for giant galaxies in CDM theory.  With the more accurate fits in equations 
20 -- 22, 
\begin{eqnarray}
  \rho_0 &\propto& r_c^{\,(-1.12 \pm 0.06)}\,;   \\
  \rho_0 &\propto& \sigma^{\,(-1.69 \pm 0.25)}\,;  \\
  \sigma &\propto& r_c^{\,(0.44 \pm 0.03)}\,.
\end{eqnarray}
we get $n = -1.80 \pm 0.10$, $n = -2.12 \pm 0.10$, and $n = -1.81 \pm 0.10$,
respectively.  Note that these values are not independent.  Their average is 
$n = -1.91$.  If we use the fits (equations 5 -- 7) determined from decompositions 
using isothermal DM, then the average is $n = -2.1 \pm 0.2$.  Both values are 
remarkably close to the value $n \simeq -2.1$ expected
in $\Lambda$CDM theory at a halo mass of $10^{12}\;M_\odot$ (Shapiro \& Iliev 
2002).  We need to correct the slopes for baryonic DM compression; this will be
done in Kormendy \& Freeman (2003).  The above comparison provides a measure 
of the slope of the fluctuation power spectrum on mass scales that are smaller
than those accessible to most other methods. 

      Shapiro \& Iliev (2002) have made a more detailed comparison of the DM
parameter correlations published by Kormendy \& Freeman (1996) with their 
predictions based on {\it COBE\/}-normalized CDM fluctuation spectra.  They found that 
the agreement between predictions and observations was best for $\Lambda$CDM.

      It is interesting to note a consequence of the theoretical prediction
that the slope $n$ gets steeper at smaller mass scales.  If $n \simeq -2.6$ 
for the smallest dwarfs (Shapiro \& Iliev 2002; Ricotti 2002), then the straight
lines in the left panels of Figure 4 should curve downward toward the visible
matter parameters of dSph galaxies.  This would strengthen the inference that the 
visible and dark matter in these galaxies is distributed similarly.  It will 
be important to look for curvature in the correlations as more data become 
available for dwarf galaxies.

      Finally, we note that the scatter in Figures 2 -- 4 has surely been 
increased by problems with the data.  (1) Distance errors are not negligible.
For our calibrating galaxies, we can compare accurate distances to those given 
by our Virgocentric flow field machinery.  This implies errors in $\log {D}$
of $\pm\,0.11$.  Since $\rho_0 \propto D^{-2}$, distance errors are a significant
-- although not the dominant -- source of scatter in equations 5 -- 25. (2) If 
some disks are submaximal, then this affects the scatter in the correlations.  
If the degree to which they are submaximal depends on $M_B$ (Kranz, Slyz, \& 
Rix 2003), this affects the correlation slopes, too.  (3) The assumption that 
DM halos have isothermal cores is challenged by CDM theory, although it is 
suppported by many observations.  It will be important to see how the 
correlations are affected if NFW halos are used. 
(4) 
The correlations in Figures 2 -- 4 require correction for DM compression by the 
baryons before a definitive comparison with theory can be made.  We will address 
these issues in future papers.

\acknowledgments

      JK is grateful to the staff of Mt.~Stromlo Observatory for their
hospitality during three visits when part of this work was done.  We thank
S.~Djorgovski, S.~M.~Fall, and P.~Shapiro for helpful discussions 
on the comparison of predicted and observed DM scaling laws.
This work used the NASA/IPAC Extragalactic Database, 
which is operated by the Jet Propulsion Laboratory, California Institute 
of Technology, under contract with NASA.

\end{document}